\documentclass[english]{article}

\usepackage{hyperref}
\usepackage[letterpaper]{geometry}
\usepackage{amsmath}
\usepackage{amssymb}
\usepackage{graphicx}
\usepackage{array}
\usepackage{tikz}
\usepackage{enumerate}
\usepackage{booktabs}
\usepackage{tabularx}
\usepackage{caption} 
\usepackage{authblk}
\usepackage{parskip}
\usepackage{bm}
\usepackage{soul}
\usepackage{float}
\usepackage{multirow}
\usepackage{wrapfig}
\newcommand{\R}{\mathbb{R}}
\DeclareMathOperator{\Tr}{Tr}

\title{Path-dependent connectivity, not modularity, consistently predicts controllability of structural brain networks}
\author[a]{Shubhankar P. Patankar}
\author[a]{Jason Z. Kim}
\author[b]{Fabio Pasqualetti}
\author[a,c,d,e,f,g,h]{Danielle S. Bassett} 

\affil[a]{Department of Bioengineering, School of Engineering and Applied Science, University of Pennsylvania, Philadelphia, PA 19104}
\affil[b]{Department of Mechanical Engineering, University of California, Riverside, CA 92521, USA}
\affil[c]{Department of Neuroscience, Perelman School of Medicine, University of Pennsylvania, Philadelphia, PA 19104}
\affil[d]{Department of Electrical and Systems Engineering, School of Engineering and Applied Science, University of Pennsylvania, Philadelphia, PA 19104}
\affil[e]{Department of Neurology, Perelman School of Medicine, University of Pennsylvania, Philadelphia, PA 19104}
\affil[f]{Department of Physics and Astronomy, College of Arts and Sciences, University of Pennsylvania, Philadelphia, PA 19104}
\affil[g]{Department of Psychiatry, Perelman School of Medicine, University of Pennsylvania, Philadelphia, PA 19104}
\affil[h]{Santa Fe Institute, Santa Fe, NM 87501}

\begin{document}
\maketitle

\begin{abstract}
The human brain displays rich communication dynamics that are thought to be particularly well-reflected in its marked community structure. Yet, the precise relationship between community structure in structural brain networks and the communication dynamics that can emerge therefrom is not well-understood. In addition to offering insight into the structure-function relationship of networked systems, such an understanding is a critical step towards the ability to manipulate the brain's large-scale dynamical activity in a targeted manner. We investigate the role of community structure in the controllability of structural brain networks. At the region level, we find that certain network measures of community structure are sometimes statistically correlated with measures of linear controllability. However, we then demonstrate that this relationship depends on the distribution of network edge weights. We highlight the complexity of the relationship between community structure and controllability by performing numerical simulations using canonical graph models with varying mesoscale architectures and edge weight distributions. Finally, we demonstrate that \emph{weighted subgraph centrality}, a measure rooted in the graph spectrum, and which captures higher-order graph architecture, is a stronger and more consistent predictor of controllability. Our study contributes to an understanding of how the brain's diverse mesoscale structure supports transient communication dynamics.
\end{abstract}

\newpage
\section{Summary}
A central question in network neuroscience is how the structure of the brain constrains the patterns of communication dynamics that underlie function. At the mesoscale of network organization, this question has been examined through the lens of modularity. Recent work has demonstrated a diversity in the mesoscale architecture of the human connectome. Further diversity in the characterization of structural brain networks is introduced by the fact that the distribution of edge weights in a network depends on the precise empirical measurement whose value is assigned to an edge. This paper explores network controllability in light of the variety of community interaction motifs and edge weight distributions that may be used to characterize structural brain networks.

\newpage
\section{Introduction}
The brain is a complex system of interconnected components that can be studied at a variety of spatial and temporal scales \cite{betzel_17}. Signals between communicating neuronal populations propagate along the white matter structure of the brain and give rise to the complex repertoire of functional dynamics that underlie cognition \cite{chialvo_10, bassett_11, fries_15, tononi_16}. A key goal of network neuroscience is to elucidate the relationship between brain network structure and function \cite{sporns_00, honey_07, honey_09,Bansal2018:DataDriven}. At any scale of interest, the patterns of inter-connectivity between components constrain the functional dynamics that may evolve on the underlying network topology \cite{wang_15}, and thus the patterns of communication between neural units. Indeed, structural brain networks display striking features such as small-worldness \cite{bassett_17}, hierarchical organization \cite{meunier_10}, spatial and topological scaling relationships \cite{bassett_10}, and modularity \cite{sporns_16}. Modularity, in particular, is a commonly studied feature of interest at the mesoscale of brain network organization that impacts potential patterns of communication.

\begin{figure}[!htbp]
\centerline{\includegraphics[width=\textwidth]{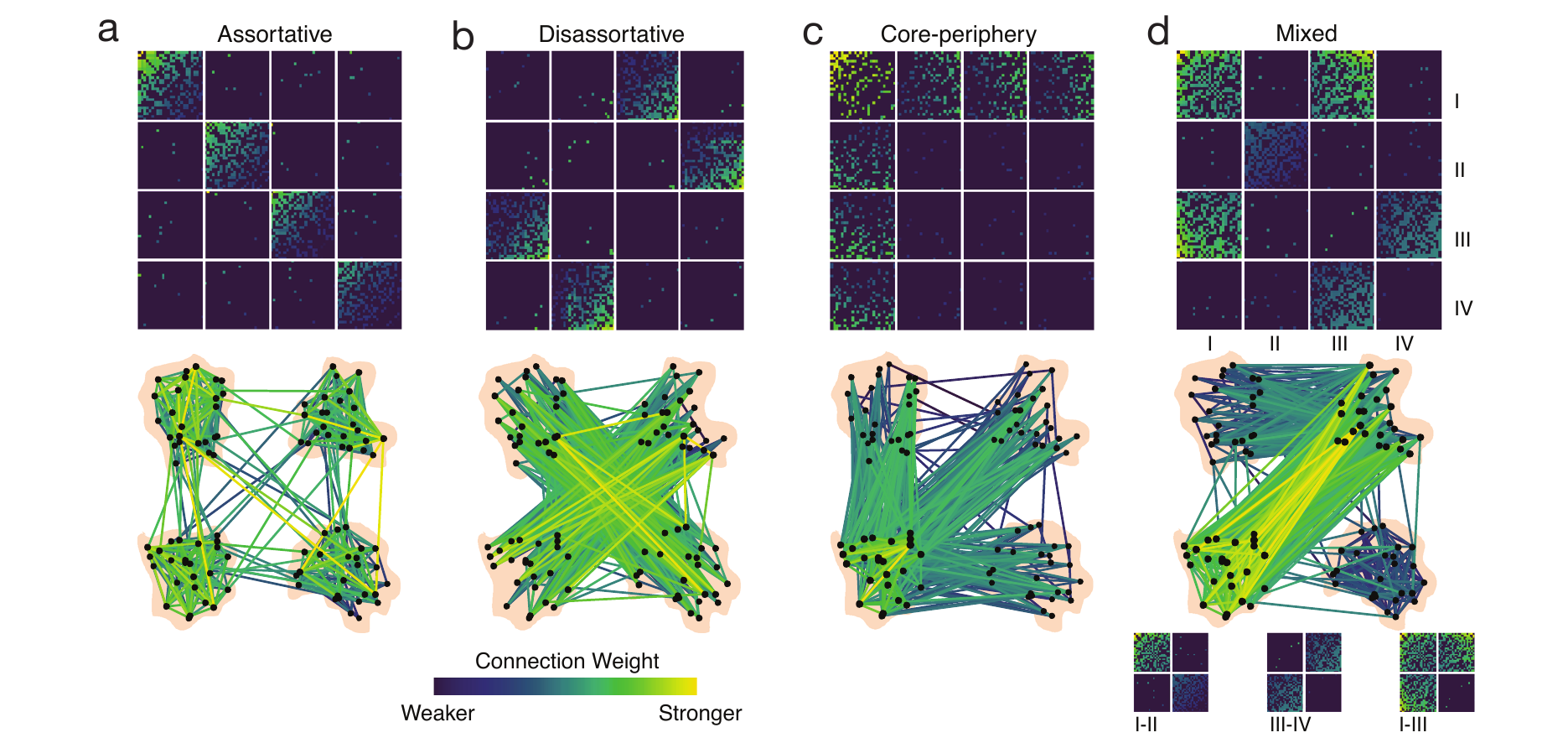}}
\caption{\textbf{Structural brain networks exhibit a diversity of mesoscale architectures.} \textbf{(a)} Assortative communities are internally densely and externally sparsely connected, whereas \textbf{(b)} disassortative communities are internally sparsely but externally densely connected. \textbf{(c)} Core-periphery organization is characterized by a dense core of well-connected nodes, and a periphery of sparsely connected nodes. \textbf{(d)} Structural brain networks have been observed to possess a mixed meso-scale architecture that combines assortative, disassortative, and core-periphery organization [Figure reproduced with permission from \cite{betzel_18}]. \label{fig:intro_fig_1}}
\end{figure}

The term ``mesoscale'' refers to the topological level higher than that of a single node, but lower than that of the entire network. Community detection techniques have been applied extensively to both structural and functional brain networks in order to group together nodes that share common features; each group is commonly referred to as a community or module. The predominant view is that the brain is composed of assortative modules, in which nodes connect densely to other nodes within their own community and sparsely to nodes outside of their community. Assortative modules are observed across species ranging from humans \cite{van_den_heuvel_11, sporns_13} and non-human primates such as macaques \cite{harriger_12}, to the nematode \textit{C. elegans} \cite{towlson_13}, and are thought to enable information integration and segregation in support of flexible cognition and behavior \cite{park_13}. However, the field's focus on assortative modules could in part be an artifact of our methodologies; popular community detection algorithms expressly seek internally dense and externally sparse sub-networks and are agnostic to other forms of mesoscale structure \cite{newman_04, newman_06, rosvall_08}. Recent work has suggested that while most brain communities are indeed assortative, others form disassortative and core-periphery structures \cite{pavlovic_14, betzel_18, faskowitz_18, faskowitz_19} (Figure \ref{fig:intro_fig_1}). The existence of such a diverse mesoscale architecture could explain the diversity of the brain's functional repertoire \cite{deco_15, betzel_18}. 

\begin{figure}[!htbp]
\centerline{\includegraphics[width=\textwidth]{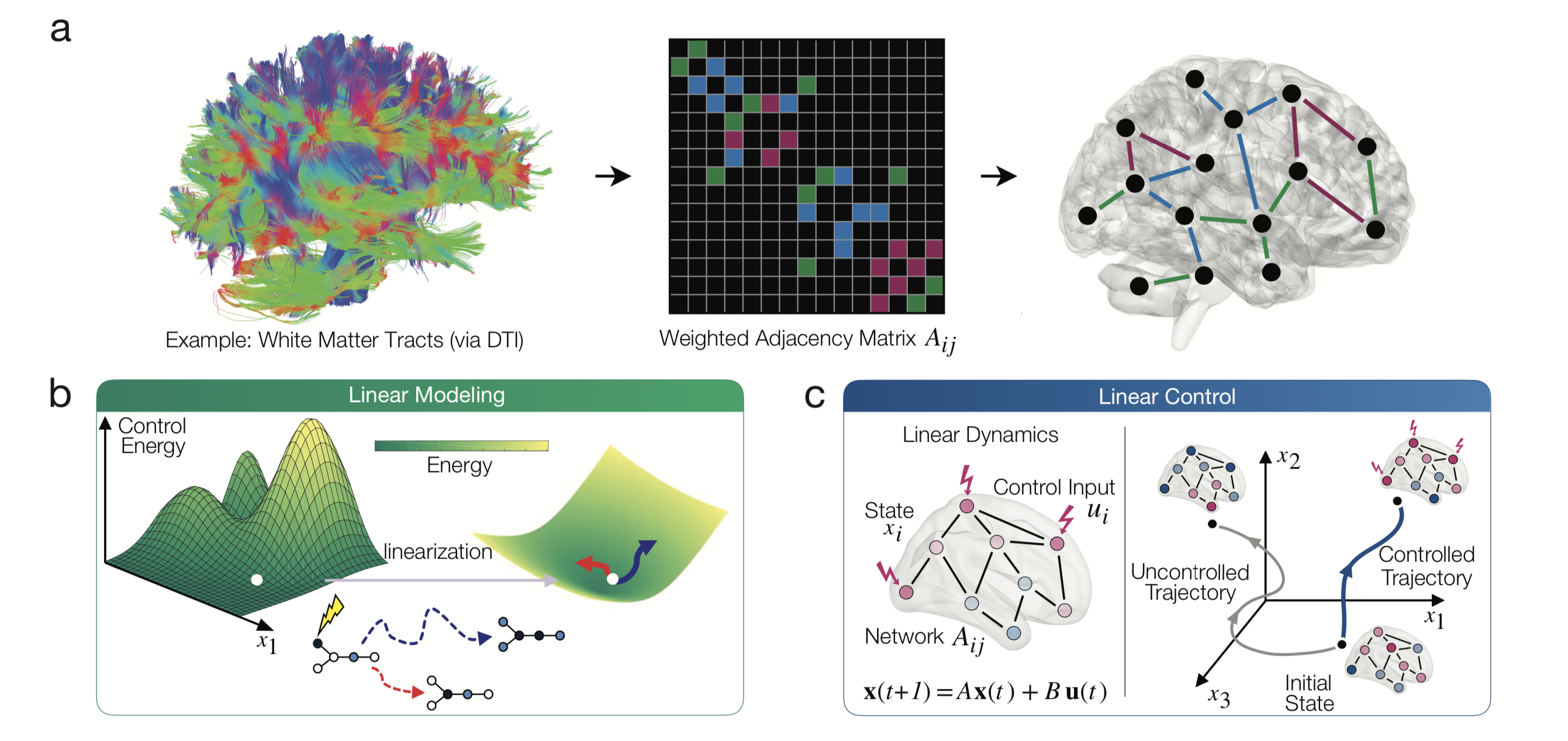}}
\caption{\textbf{Schematic of methods and approach.} (\textbf{a}) A variety of empirical measurements are used to estimate and study brain network structure. This data is then compiled into a weighted network adjacency matrix $A$ whose entries $A_{ij}$ describe the connection strength of region $i$ and region $j$, thus characterizing the brain's structural network. \textbf{(b)} While brain dynamics are non-linear, linearization is a convenient modeling approach that has been demonstrated to yield biologically meaningful insights, and one that allows us to systematically investigate relationships between model parameters and model behavior. Linear systems theory provides a natural language in which to characterize state transitions in the brain. (\textbf{c}) The level of activity in each brain region is combined into a state vector $\bm{x}$ and modeled using a linear dynamical system. Linear control theory can be used to assess the effect of exogenous inputs on the brain's functional dynamics. Controllability may be quantified using metrics such as average and modal controllability, and the minimum energy required to cause a state transition [Figure reproduced with permission from \cite{lynn_19}]. \label{fig:intro_fig_2}}
\end{figure}

Yet, precisely how the community structure of brain networks constrains, supports, and explicates the communication dynamics that we observe in empirical measurements is not well understood. Whole-brain models of neural dynamics provide an avenue to bridge this knowledge gap by stipulating how neural activity propagates along the underlying structural network \cite{avena_17, lynn_19}. Further insight into how transient dynamics evolve on networks can be obtained by perturbing the dynamical model with exogenous inputs. Linear systems theory and its associated network control framework can be used to probe the relationship between the structure of networks and the transient dynamics that they support \cite{kailath, liu_11} (Figure \ref{fig:intro_fig_2}b). The approach requires that the brain be represented as a network of regions connected by edges, which are commonly derived from empirical estimates reflecting the strength, volume, or integrity of white matter tracts \cite{bassett_17_net_neur,bassett2018on} (Figure \ref{fig:intro_fig_2}a). Control inputs, which are representative of changing levels of activity, can then be added to network nodes to study the evolution of activity dynamics \cite{gu_15, tang_18} (Figure \ref{fig:intro_fig_2}c). From a biophysical perspective, these inputs may represent an endogenous shift in neural activity from one cognitive state to another \cite{gu_15, cornblath_19}, or even direct exogenous inputs such as during electrical stimulation \cite{stiso_19,khambhati2019functional}. 

We hypothesize that brain regions have different controllability statistics depending on the extent to which they participate in interactions with nodes from other communities. We reason that a diversity in connections ought to lead to greater ability for a node to control the rest of the network. To test this hypothesis, we partition brain regions into communities by applying the weighted stochastic block model (WSBM) to structural connectivity matrices extracted from non-invasive magnetic resonance imaging (MRI) measurements in humans. Block modeling is a flexible community detection technique that is able to uncover diverse mesoscale motifs beyond the commonly studied assortative type \cite{hastings_06, aicher_15}. The connectivity matrices we study encode networks whose nodes represent brain regions. Edges can represent diverse estimates of inter-node connections, such as white matter streamline counts between regions, mean quantitative anisotropy (QA) values along the streamlines, and generalized fractional anisotropy values (GFA) \cite{tuch_04, hagmann_07, smith_12, yeh_13}. Unfortunately, there is no consensus in the field yet regarding whether one type of edge weight has more utility than another type of edge weight, and therefore the literature contains studies that use a variety. The distribution of edge weights in the network depends on the precise quantity that the edge represents, and this fact hampers formal comparison of results across studies. For example, structural brain networks with QA values \cite{kim_18, stiso_19} and those with streamline counts have differing edge weight distributions. Both have been previously used for network control theoretic studies \cite{stiso_19, kim_18, karrer_20, gu_15, cornblath_19,lee2019heritability,jeganathan_18,shine_19}, but direct comparisons between the two have not been performed. Here we seek to obtain a more comprehensive understanding of the relations between community structure and controllability that is independent of the choice of edge weight, and the associated differences in edge weight distribution. Thus, we use multiple data sets containing networks with distinct edge definitions. 

We further hypothesize that disrupting the amount of a particular mesoscale motif such as assortativity, disassortativity, or core-peripheriness in a network ought to result in a motif-specific controllability profile. We perform numerical simulations to gradually alter the mesoscale structure of networks along specific continuums of interest while preserving their binary density and the distribution from which network edge weights are drawn. At each stage, we examine their controllability. In one set of simulations we alter the binary topology on an axis ranging from disassortative to assortative. In another set of simulations, network topology ranges from disassortative to core-periphery. We perform both sets of simulations on networks where edge weights are drawn from the normal distribution as well as the geometric distribution. The latter distribution is an example of a fat-tailed distribution, which resembles the weighted degree distributions of many biologically observed networks \cite{broido_19}. If binary topology of networks is the key driver of controllability, we expect to observe that regardless of the choice of distribution used to assign edge weights; similar alterations to network topology along a structural continuum ought to similarly affect patterns of network controllability.

\newpage
\section{Mathematical Framework}
While brain network dynamics are known to be nonlinear (Figure \ref{fig:intro_fig_2}b) \cite{rabinovich_06}, the simplification to a linearized network model is often a useful approximation \cite{galan_08, ABDELNOUR2014335}. We offer a discussion of the utility of the linear framework in the `Discussion' section; for a more comprehensive discussion we point the reader to the Supplement.

A linear model may be created by linearizing the non-linear system of interest about a fixed point. System dynamics are then characterized in terms of deviations about this fixed point. Linear modeling provides a tractable simplification for the analysis of non-linear dynamical systems, allowing the use of well-developed theoretical tools from linear systems and control theory to investigate network dynamics in response to exogenous control inputs \cite{kailath}. In the context of brain networks, the linear model allows one to study how signals can propagate along structural links connecting brain regions. 

Suppose we have a node set $\mathcal{V} = \{1,\cdots,n\}$ with undirected weighted edges $\mathcal{E} \subseteq \mathcal{V} \times \mathcal{V}$, compiled in a graph $\mathcal{G} = (\mathcal{V},\mathcal{E})$ and represented by a symmetric weighted adjacency matrix $A \in \R^{n \times n}$. Elements of $\mathcal{V}$ denote brain regions and elements of $\mathcal{E}$ represent the strengths of the connection between them. The dynamics of a discrete-time linear time-invariant LTI system are written as
\begin{equation}
    \bm{{x}}(t+1) = A\bm{x}(t) + B\bm{u}(t),
\end{equation}
where $A$ is the $n \times n$ symmetric and weighted network adjacency matrix, which acts as the system matrix in the LTI framework, and $B$ is an $n \times k$ matrix, where $k$ is the number of independent control inputs.
A full control set implies that all $n$ network nodes receive input, for instance in the case when $B = I_n$, the identity matrix of dimension $n$. The terms $\bm{x}(t)$ and $\bm{u}(t)$ represent the state of the system and the exogenous input at time $t$, respectively (see `Discussion' for biophysical interpretations of $\bm{x}(t)$ and $\bm{u}(t)$).

A particularly useful element of the linear control framework is the matrix defined as, 
\begin{equation}\label{eq:gramian}
    {W_C}(T) = \sum_{t = 0}^{T - 1} A^tBB^{\top}(A^{\top})^{t}
\end{equation}
called the \textit{finite time controllability Gramian}, where $T$ refers to the time horizon of control \cite{kailath}. The Gramian plays a vital role in determining the unique control input of minimum energy that transitions the network state from some initial state $\bm{x}_0$ at $t = 0$ to a final state $\bm{x}_f$ at a later time $t = T$ \cite{stiso_19, karrer_20}. We create target state vectors by placing a 1 in $\bm{x}_f$ corresponding to the location of each brain region $i$ in turn, and 0s elsewhere. These one-hot vectors may be thought to represent the activation of a single brain region with a full control set.
With $\bm{x}_0 = \mathbf{0}$, the minimum energy of the input required to attain a state $\bm{x}_f$ at time $T$ is written as,
\begin{equation}
    E_{i} = {\bm{x}_f}^{\top} {W_C}^{-1}(T) {\bm{x}_f}.
\end{equation}
We demonstrate in the Supplement that the energies thus computed, by performing $N$ state transitions to $N$ one-hot vectors, form an upper bound on the energy required to perform arbitrary non-negative state transitions.

In addition to the useful energy-related interpretation, other controllability metrics are often defined using the Gramian \cite{FP-SZ-FB:13q}. Average controllability, which is the average energy input over all possible target states \cite{marx_04, shaker_12}, is one such metric. It has been used in previous studies examining the controllability of structural brain networks \cite{jeganathan_18, bernhardt_19, lee_19,lee2019heritability, shine_19}. Average controllability is proportional to the trace of the inverse of the controllability Gramian, $\Tr(W_C^{-1})$. In practice however, this quantity is replaced by the trace of the controllability Gramian, $\Tr(W_C)$, since computing the inverse of $W_C$ is typically ill-conditioned, and the two quantities satisfy a bounded relation of inverse proportionality \cite{FP-SZ-FB:13q, summers_14}. We compute average controllability for an individual node by setting $B = b_i$, where $b_i$ is a one-hot vector with a $1$ in the location corresponding to a node. Smaller values of average controllability for a node may be thought of as implying that the network is less controllable on average from that node. 

Another controllability measure that is often used in the context of structural brain networks is modal controllability \cite{FP-SZ-FB:13q, stiso_19, gu_15, karrer_20,khambhati2019functional,shine_19}. Modal controllability quantifies the extent to which a network's eigenmodes, weighted by the rate of their decay, are influenced by input into a brain region. For a node $i$, modal controllability is defined as: $\phi_{i}=\sum_{j=1}^{N}\left(1-\lambda_{j}^{2}(A)\right) v_{i j}^{2}$ \cite{karrer_20}. We note that this functional form of modal controllability is defined specifically for symmetric matrices. Here, $\lambda_j$ represents an eigenvalue of the weighted adjacency matrix and $v_{ij}$ represents the $i$-th component of the $j$-th eigenvector of $A$. Since the weighted adjacency matrix is symmetric, all of its eigenvalues are real. The eigenvectors of $A$ represent independent directions in the state-space along which system dynamics evolve according to the rate specified by the corresponding eigenvalues. A quickly decaying mode is harder to control since, intuitively, it requires more input energy to sustain its activity. As a result, this metric has been previously described as being a measure of the controllability to the `hard-to-reach' states of a system \cite{gu_15, tang_17, cornblath_19}. 

In order to ensure comparability of time scales across networks, we scale the network adjacency matrices by their largest eigenvalues. In this study we set $T = 4$ for average controllability and minimum energy computations. However, we demonstrate that our results remain robust to a broad range of choices of $T$ in the Supplement. We also note that whereas average/modal controllability consider control from a single node, minimum control energy considers controllability from a larger node set. All minimum control energy results presented in this paper are computed using a full control set, $B = I$.

\section{Results}

\subsection{Relationship between network controllability and community structure for edge weights drawn from a normal distribution}
Results presented in this section are obtained from analyses performed on Data Set 1 (see subsection `Data' in the `Methods' for details), which is comprised of structural brain networks where edges represent estimates of mean quantitative anisotropy (QA) values. An element $[A_{ij}]$ of the weighted adjacency matrix for these networks represents the mean QA weighting across streamlines connecting two regions $i$ and $j$. Note that edge weights with QA values approximate a normal distribution. 

\subsubsection{Measures of controllability are not consistently correlated with measures of modularity for structural brain networks with normally distributed edge weights}

Prior work has reported a statistical correlation between some controllability metrics and modularity, a summary measure of assortative community structure \cite{tang_17}; yet, importantly in that study results held even after regressing out the effects of modularity. Here we began our investigation by assessing whether controllability of structural brain networks is statistically related to community structure in a different data set than the one used by Tang \emph{et al.}, and when using a larger set of measures of a network's community structure. Specifically, we compute three metrics of network control for each brain region: minimum control energy to activate the region, average controllability, and modal controllability. We then study the relationships between these measures, and the weighted variant of the participation coefficient and the intra-module strength $Z$-score. Participation coefficient measures the diversity of the distribution of a node's strength amongst network modules. A value of $0$ for a node implies that all its connection strength is associated with other nodes in its own module, whereas a value of $1$ implies that connection strength is distributed uniformly among all modules. Intra-module strength $Z$-score measures the connectivity strength of a node to other nodes in its own module \cite{guimera_05, rubinov_11}. We compute participation coefficient for brain regions and the intra-module strength $Z$-score after partitioning the networks into communities using the weighted stochastic block model (WSBM). We use the normal distribution as the choice of prior for the edge weight distribution when applying the WSBM, since edge weights in QA weighted networks are approximately normally distributed. 

We begin by testing the relationships between participation coefficient and the intra-module strength $Z$-score, and the three measures of network controllability. We observe that participation coefficient relates negatively with minimum control energy ($\rho = -0.807$, $p \approx 0$) and with modal controllability ($\rho = -0.810$, $p \approx 0$), whereas it relates positively with average controllability {($\rho = 0.815$, $p \approx 0$)}. Similarly, intra-module strength $Z$-score relates negatively with both minimum control energy ($\rho = -0.338$, $p \approx 0$) and modal controllability {($\rho = -0.323$, $p \approx 0$)}, and relates positively with average controllability ($\rho = 0.244$, $p \approx 0$). These observations suggest the presence of a statistical relationship between community structure and controllability. 

However, it is possible for community structure and controllability to be related due the influence of a third variable. We hypothesize that node strength could be such a shared driver since prior work has reported a correlation between network controllability and node strength \cite{gu_15, muldoon_16, lee2019heritability, jeganathan_18}. In this dataset, node strength relates negatively with minimum control energy ($\rho = -0.998$, $p \approx 0$) and with modal controllability ($\rho = -0.998$, $p \approx 0$), whereas it relates positively with average controllability ($\rho = 0.986$, $p \approx 0$). Further, we find that node strength is also positively related to both participation coefficient ($\rho = 0.807$, $p \approx 0$) and intra-module strength $Z$-score ($\rho = 0.333$, $p \approx 0$). As a result, node strength may be the potential driver of any relationship between community structure and controllability. 

\begin{figure}[!htbp]
\centerline{\includegraphics[width=\textwidth]{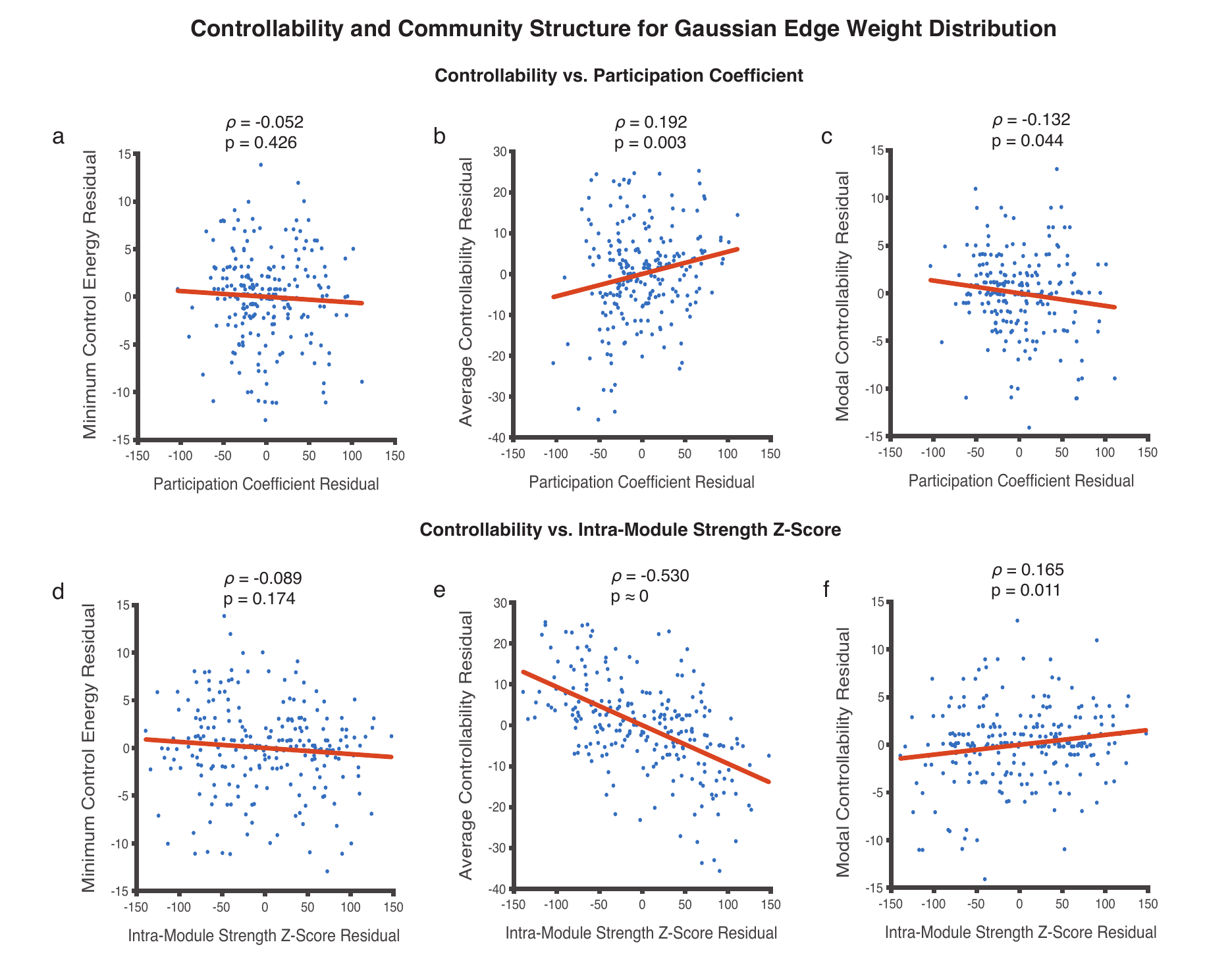}}
\caption{\textbf{Relationships between metrics of regional controllability and metrics of community structure for edge weights approximating a normal distribution.} (\textbf{a, b, c}) Participation coefficient does not relate in a statistically significant manner with minimum control energy ($\rho = -0.052$, $p = 0.426$) when accounting for node strength. On the other hand, correlations between participation coefficient with average ($\rho = 0.192$, $p = 0.003$) and modal controllability ($\rho = -0.132$, $p = 0.044$) survive corrections for node strength. (\textbf{d, e, f}) Intra-module strength $Z$-score follows a similar pattern; it does not relate with minimum control energy ($\rho = -0.089$, $p = 0.174$), but relates significantly with average ($\rho = -0.530$, $p \approx 0$) and modal controllability ($\rho = 0.165$, $p = 0.011$). Each dot in the scatter plots represents the mean value of a controllability and modularity measure across 24 (8 subjects in triplicate) network instantiations for a single brain region resulting in 234 data points. \label{fig:corr_gauss}}
\end{figure}

Therefore, we run partial Spearman correlations between metrics of community structure and controllability, correcting for node strength (Figure \ref{fig:corr_gauss}). We find that when node strength is accounted for, participation coefficient no longer relates to minimum control energy ($\rho = -0.052$, $p = 0.426$) (Figure \ref{fig:corr_gauss}a). It continues to relate significantly with average controllability ($\rho = 0.192$, $p = 0.003$) and modal controllability ($\rho = -0.132$, $p = 0.044$) (Figure \ref{fig:corr_gauss}b, c). Intra-module strength $Z$-score follows a similar trend; it does not relate significantly with minimum control energy ($\rho = -0.089$, $p = 0.174$), but continues to relate with average controllability ($\rho = -0.530$, $p \approx 0$) and modal controllability ($\rho = 0.165$, $p = 0.011$) even when controlling for node strength (Figure \ref{fig:corr_gauss}d, e, f).

From the findings in this section, we conclude that for the examined structural brain networks where edge weights are approximately normally distributed, region-level measures of modularity such as participation coefficient and intra-module strength $Z$-score correlate in a statistically significant manner with average and modal controllability, but not with minimum control energy.

\subsubsection{Numerical simulations using edges drawn from a normal distribution}
Next, we seek to better understand the relationship between controllability and community structure by parsing community structure into distinct motifs, such as assortativity, or core-peripheriness. We generate synthetic networks with a specifically determined community structure and examine their controllability. \textit{In silico} experiments where network topologies are precisely enforced and edge weights are drawn from distributions with precisely known parameters are useful benchmarks in understanding the relationship between mesoscale organization and controllability. We begin by generating networks with a $2 \times 2$ block structure in their adjacency matrices, and with normally distributed edge weights (see subsection `Numerical Simulations' in the `Methods' for details). 

Recall that when the diagonal blocks of a network are denser relative to the off-diagonal blocks, networks possess an assortative block structure (Figure \ref{fig:intro_fig_1}a). By contrast, when the off-diagonal blocks are denser relative to the diagonal blocks, network communities interact disassortatively (Figure \ref{fig:intro_fig_1}b). Another form of mesoscale topology is the core-periphery structure (Figure \ref{fig:intro_fig_1}c). Nodes in the core are connected more densely to each other than they are to the rest of the network. Nodes in the periphery predominantly connect with nodes in the core but not with each other. We quantify the notion of modularity in the form of the modularity quality index ($Q$), which is a network-level measure of how well a given community partition segregates nodes into modules. It quantifies the extent of modularity by relating the observed strength of within-module connections in a network to the strength of within-module connections expected under a null model \cite{newman_04}. The quantity $Q$ can be positive or negative, with positive values implying the presence of an assortative community structure \cite{newman_06}. We characterize the relationship between $Q$ and the fraction of network edges inside of modules (or the core) in the Supplement.

\begin{figure}[!htbp]
\centerline{\includegraphics[width=\textwidth]{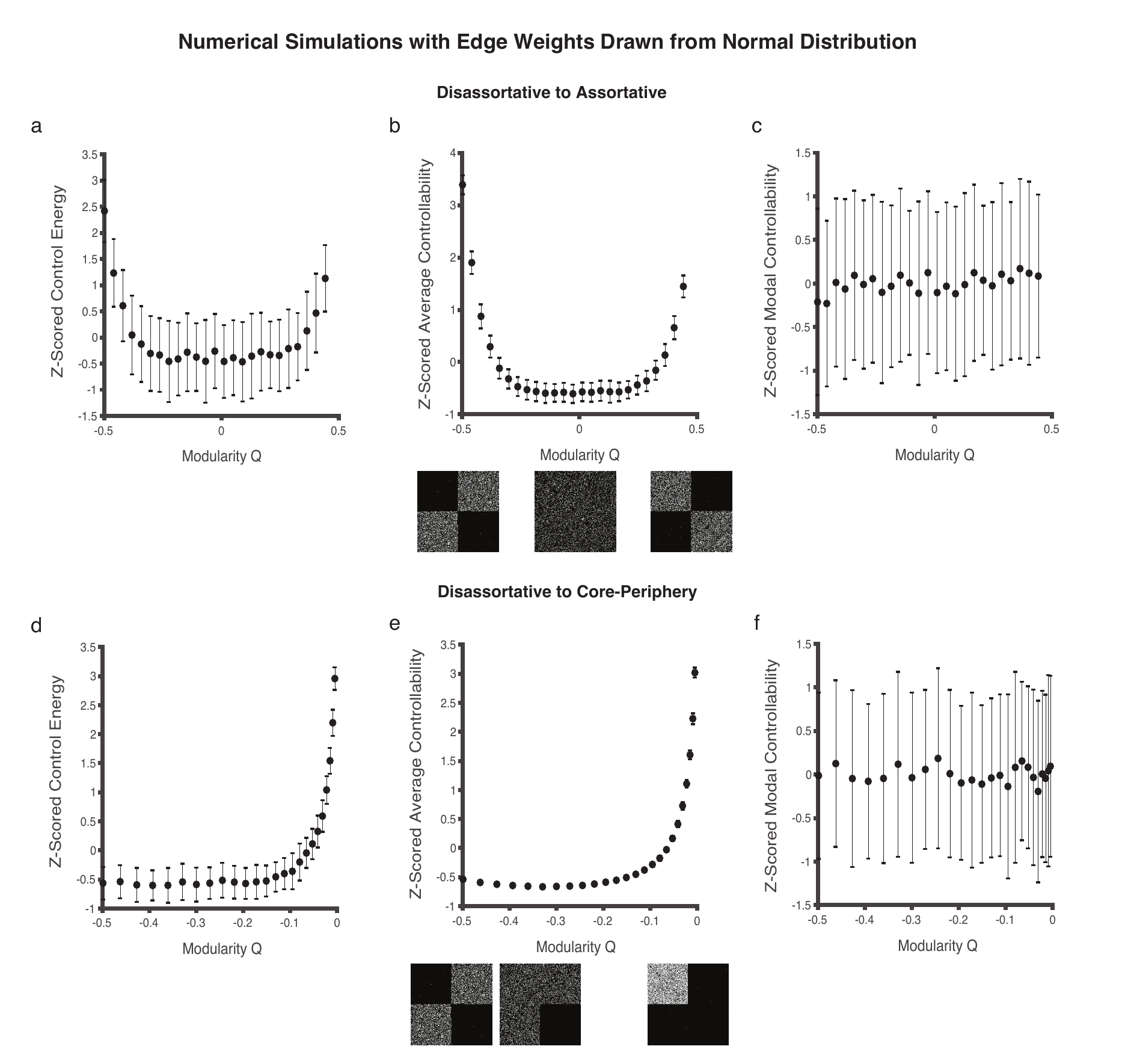}}
\caption{\textbf{Controllability for normally weighted networks as a function of changing mesoscale topology.} (\textbf{a, b}) As network topology changes from disassortative to assortative, mean network control energy and average controllability first decrease, and then increase tracing out U-shaped curves. Their values are the lowest when $Q \approx 0$, which corresponds to the point of randomness. Networks with a balance between disassortativity and coreness occur when $Q \approx -0.28$. (\textbf{d}) Minimum control energy increases as networks become less disassortative and more core-like. (\textbf{e}) Average controllability first decreases and then rapidly increases past $Q \approx -0.28$. (\textbf{c, f}) Modal controllability, on the other hand, exhibits no discernible trends with changing network topology. Each point in the scatter plots represents a $Z$-scored mean network controllability value computed across $100$ network instantiations at each $Q$-value. Error bars correspond to the standard deviation of the mean controllability value for networks in a given ensemble. \label{fig:sims_gauss}}
\end{figure}

In the first set of simulations, we generate networks on a range from disassortative to assortative (see subsection `Numerical Simulations' in the `Methods' for details). At each point along the structural continuum, we generate an ensemble of $100$ different sparse weighted networks with a known value of the modularity quality index $Q$. First, for each network in the ensemble we compute the mean of the 234 obtained values of minimum control energy, average controllability, and modal controllability. Minimum control energy and average controllability values are computed using $T = 4$ as the choice of time horizon for consistency. We then compute the mean of the three network-level controllability metrics across the $100$ network instantiations in the ensemble. We observe that as network topology becomes more assortative from disassortative, minimum control energy and average controllability first decrease, and then increase with a minimum value at $Q \approx 0$ (Figure \ref{fig:sims_gauss}a, b). The trough corresponds to $Q \approx 0$ where the network topology is random. Modal controllability has no discernible trend with changing network topology along the disassortative-assortative continuum (Figure \ref{fig:sims_gauss}c).

In the second set of simulations, we generate networks on a range from disassortative to core (see subsection `Numerical Simulations' in the `Methods' for details). Along this structural continuum, when the fraction of edges in the core ($[1,1]$-block) is closer to $0$, a network is disassortative, whereas when the fraction is closer to $1$, it has a dense core reminiscent of a core-periphery network. Networks are nearly random when the fraction is $1/3$ for the $2 \times 2$ block adjacency matrix with a single on-diagonal block ($[2,2]$-block) having zero density. In terms of the modularity quality index $Q$, the extremes correspond to values of $-0.5$ (disassortative) and $0$ (core), respectively. The extent of disassortativity and coreness is in balance when $Q \approx -0.28$. Similar to the first set of simulations, we generate $100$ network instantiations as the topology gradually changes from disassortative to more core-like. We observe that as networks become more core-like, mean minimum control energy increases (Figure \ref{fig:sims_gauss}d). There is little change in the mean control energy value in the disassortative regime; however, this is followed by a sharp rise past $Q \approx -0.20$. Average controllability, in contrast, first decreases gradually to $Q \approx -0.28$, followed by a sharp increase (Figure \ref{fig:sims_gauss}e). Similar to the disassortative-assortative structural continuum, modal controllability does not exhibit a significant trend along the disassortative-core continuum (Figure \ref{fig:sims_gauss}f).

In summary, disruptions to particular mesoscale motifs in networks where edges are drawn from a normal distribution result in motif-specific profiles of network controllability.

\subsection{Relationship between network controllability and community structure for edge weights drawn from a fat-tailed distribution}

In the context of structural brain networks, multiple empirical estimates may be used to quantify the strength of connections between two regions, such as white matter streamline counts between regions, mean quantitative anisotropy (QA) values along the streamlines, and generalized fractional anisotropy (GFA) values. These measures reflect the strength, volume, or integrity of white matter tracts connecting one region of the brain to another. This diversity in the characterization of structural networks introduces further complexity in the modeling of large-scale communication dynamics in the brain. The distribution of edge weights in a structural brain network is contingent on the choice of edge definition, which has the potential to cause conflict in results that relate network topology to controllability. 

In order to examine the relationship between the edge weight distribution that underlies a mesoscale topology and network controllability, we next turn to brain networks with an edge weight distribution distinct from the already examined normal distribution from Data Set 1. Results presented in this section are obtained from analyses performed on Data Set 2 (see subsection `Data' in the `Methods' for details), which is comprised of structural brain networks where edges represent estimates of streamline counts between regions. An element $[A_{ij}]$ of an adjacency matrix for these networks represents the number of streamlines connecting two brain regions $i$ and $j$. Edge weights with streamline counts approximate a fat-tailed distribution. Recent work has indicated that real-world networks with fat-tailed distributions can often be approximated using the log-normal distribution \cite{broido_19}. As a result, we use the log-normal distribution as the choice of edge weight distribution prior when inferring communities using the weighted stochastic block model (WSBM). We demonstrate the robustness of our results to the choice of the edge weight distribution prior in the Supplement.

\subsubsection{Measures of controllability are not consistently correlated with measures of modularity for structural brain networks with a fat-tailed distribution of edge weights}

Similar to our observations in structural brain networks with normally distributed edge weights (Data Set 1), here we find that the participation coefficient relates negatively with minimum control energy ($\rho = -0.433$, $p \approx 0$) and with modal controllability ($\rho = -0.435$, $p \approx 0$), and positively with average controllability ($\rho = 0.450$, $p \approx 0$) for networks with a fat-tailed edge weight distribution (Data Set 2). Intra-module strength $Z$-score relates negatively with both minimum control energy ($\rho = -0.638$, $p \approx 0$) and modal controllability ($\rho = -0.630$, $p \approx 0$), and relates positively with average controllability ($\rho = 0.565$, $p \approx 0$). These observations, yet again, suggest the presence of a statistical relationship between community structure and controllability.

\begin{figure}[!htbp]
\centerline{\includegraphics[width=\textwidth]{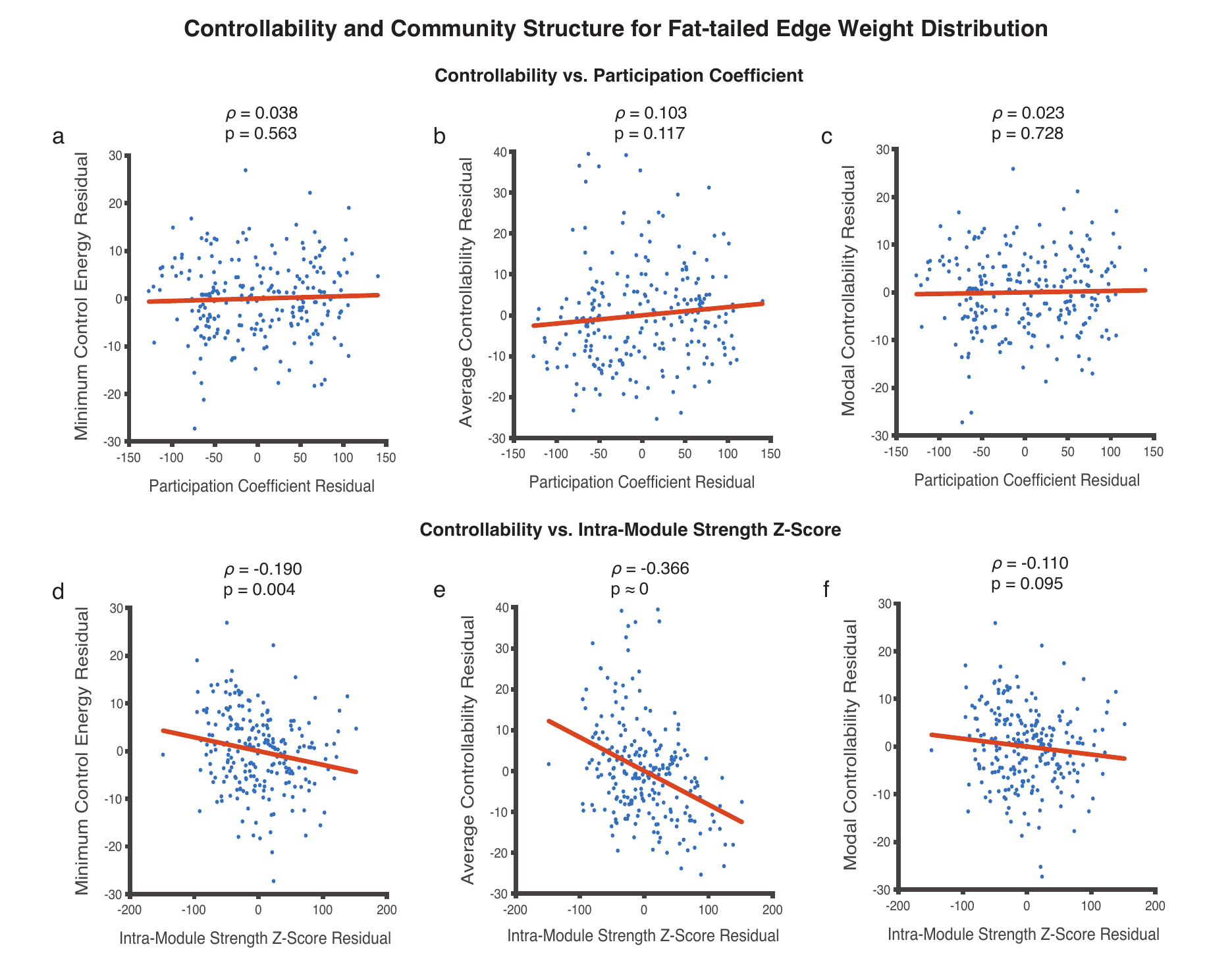}}
\caption{\textbf{Relationships between metrics of regional controllability and metrics of community structure for edge weights approximating a fat-tailed distribution.} (\textbf{a, b, c}) Participation coefficient does not relate in a statistically significant manner with minimum control energy ($\rho = 0.038$, $p = 0.563$), average controllability ($\rho = 0.103$, $p = 0.117$), or modal controllability ($\rho = 0.023$, $p = 0.728$). (\textbf{d, e}) Intra-module strength $Z$-score relates significantly with minimum control energy ($\rho = -0.190$, $p = 0.004$) and average controllability ($\rho = -0.366$, $p \approx 0$). (\textbf{f}) It does not relate with modal controllability ($\rho = -0.110$, $p = 0.095$). Each point in the scatter plots represents the mean value of a controllability and modularity measure across 24 (8 subjects in triplicate) network instantiations for a single brain region resulting in 234 data points. \label{fig:corr_stream}}
\end{figure}

Similar to Data Set 1, however, it is possible for these statistical relations between controllability and community structure to be driven by a third variable such as node strength. Indeed in Data Set 2, we also observe that node strength is related to measures of network controllability. Node strength relates negatively with minimum control energy ($\rho = -0.993$, $p \approx 0$) and modal controllability ($\rho = -0.993$, $p \approx 0$), and relates positively with average controllability ($\rho = 0.984$, $p \approx 0$). Node strength is also a predictor of the participation coefficient ($\rho = 0.440$, $p \approx 0$) and the intra-module strength $Z$-score ($\rho = 0.625$, $p \approx 0$). Similar to earlier analyses, we run partial Spearman correlations in order to account for the effects of node strength when characterizing the relationship between measures of controllability and those of community structure. We find that participation coefficient no longer significantly relates to minimum control energy ($\rho = 0.038$, $p = 0.563$) (Figure \ref{fig:corr_stream}a), average controllability ($\rho = 0.103$, $p = 0.117$) (Figure \ref{fig:corr_stream}b), or modal controllability ($\rho = 0.023$, $p = 0.728$) (Figure \ref{fig:corr_stream}c). Intra-module strength $Z$-score continues to relate in a statistically significant manner with minimum control energy ($\rho = -0.190$, $p = 0.004$) (Figure \ref{fig:corr_stream}d) and average controllability ($\rho = -0.366$, $p \approx 0$) (Figure \ref{fig:corr_stream}e), but not with modal controllability ($\rho = -0.110$, $p = 0.095$) (Figure \ref{fig:corr_stream}f) when accounting for the effect of node strength.

From the findings in this section, we conclude that for structural brain networks with a fat-tailed edge weight distribution, region-level minimum control energy and average controllability are related in a statistically significant manner with intra-module strength $Z$-score. However, unlike Data Set 1 no measure of controllability relates with participation coefficient in a statistically significant manner. Therefore, the hypothesized relationship between a node's participation in the community structure, and its associated controllability metrics, is not general and is also strongly contingent on the distribution from which network edges are drawn.

\subsubsection{Numerical simulations using edges drawn from a geometric distribution}
In parallel to the previous set of numerical simulations on networks with normally distributed edge weights, we next sought to describe the relationship between mesoscale architecture and network controllability for networks with a fat-tailed edge weight distribution. We use the geometric distribution as a representative fat-tailed distribution when drawing network edge weights.

\begin{figure}[!htbp]
\centerline{\includegraphics[width=\textwidth]{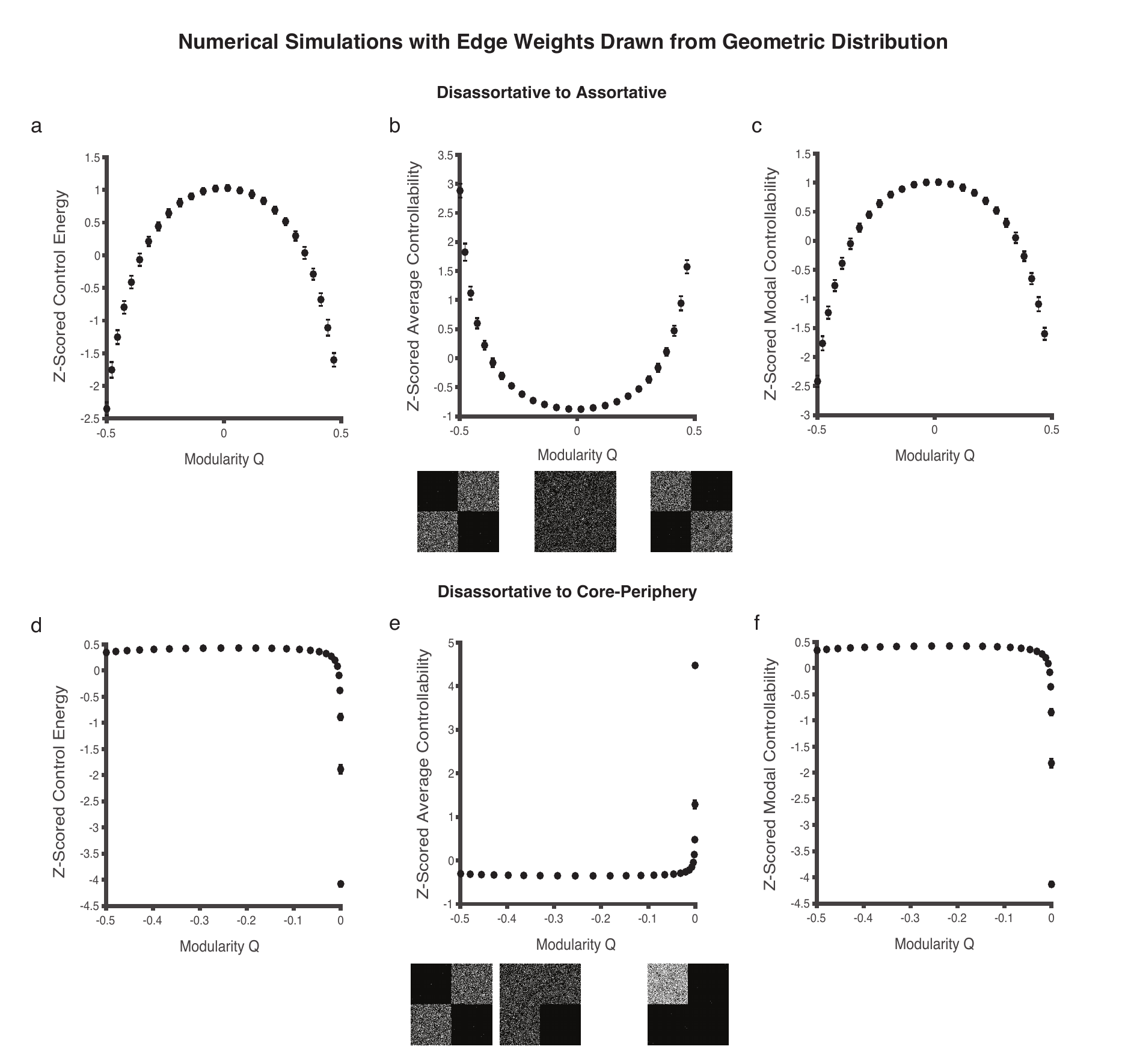}}
\caption{\textbf{Controllability for weighted networks with a geometric distribution of edge weights as a function of changing mesoscale topology.} (\textbf{a, c}) As network topology changes from disassortative to assortative, the mean network control energy and modal controllability first increase and then decrease on either side of $Q \approx 0$, which marks the point of randomness. (\textbf{b}) By contrast, average controllability exhibits the opposite trend; first decreasing and then increasing as networks become more assortative from disassortative. (\textbf{d, f}) Along the continuum from disassortativity to coreness, minimum control energy and modal controllability decrease, whereas (\textbf{e}) average controllability increases. Each point in the scatter plots represents a $Z$-scored mean network controllability value computed across $100$ network instantiations. Error bars correspond to the standard deviation of the mean controllability value for networks in a given ensemble.\label{fig:sims_geom}}
\end{figure}

In the first set of simulations, we generate networks on a range from disassortative to assortative. At each value of the modularity quality index $Q$, we generate an ensemble of $100$ sparse weighted networks with edge weights drawn from the geometric distribution (see subsection `Numerical Simulations' in the `Methods' for details). We begin by computing the mean of the nodal values of minimum control energy, average controllability, and modal controllability. We then compute the mean of the three controllability measures across the $100$ instantiations in an ensemble, and repeat this process at every $Q$ value.

We observe that as the network topology becomes more assortative from disassortative, minimum control energy and modal controllability first increase, and then decrease with a peak at $Q \approx 0$, which corresponds to the point of randomness (Figure \ref{fig:sims_geom}a, c). Average controllability, on the other hand, follows the opposite trend, and is the highest at points of greatest disassortativity and assortativity, with a low at $Q \approx 0$ (Figure \ref{fig:sims_geom}b). Importantly, the trends in network controllability observed for networks with a fat-tailed distribution (Figure \ref{fig:sims_geom}) of edge weights are not similar to those observed for networks with a normal distribution of edge weights (Figure \ref{fig:sims_gauss}).

In the second set of simulations, we generate networks on a range from disassortative to core-like (see subsection `Numerical Simulations' in the `Methods' for details). Along this structural continuum, when the modularity quality $Q$ index is closer to $-0.5$, a network is disassortative, whereas when the index is closer to $0$, it has a dense core reminiscent of a core-periphery network. Networks are nearly random when the index is $-0.28$. We find that networks with increasingly dense cores have lower mean minimum control energy and mean modal controllability (Figure \ref{fig:sims_geom}d, f). Average controllability, in contrast, increases with an increasingly dense core (Figure \ref{fig:sims_geom}e). Trends in the mean network controllability values along the disassortative-core continuum appear to form traces of U-shaped curves.

For networks where edge weights are drawn from the geometric distribution, disruptions to particular mesoscale motifs results in motif-specific profiles of network controllability. However, these profiles are distinct from those observed for networks with normally distributed edge weights. Had binary topology been a unique predictor of network controllability, the trends in the curves in Figures \ref{fig:sims_gauss} and \ref{fig:sims_geom} would have been similar for similarly altered networks along the continuums.

\subsection{Weighted subgraph centrality as a predictor of network controllability}
Based on the results thus far, and contrary to the initial hypothesis, the extent of a node's participation in the network's community structure is not a consistent predictor of its metrics of controllability. In addition, at the network-level, binary topology does not uniquely determine controllability.  It is apparent that the distribution of edge weights is as important to network controllability as the binary distribution of edges themselves. Since modularity and controllability do not uniquely explain one another, perhaps a different but complementary feature of network organization relates the two. Since eigenvalues and eigenvectors fully and uniquely describe a matrix, the spectrum of the weighted network adjacency matrix, which acts as the system matrix $A$ for our discrete-time LTI system, encodes all features of the network including those that consistently predict controllability. Therefore, we hypothesize that a node-level metric that is rooted in the graph spectrum ought to relate to controllability statistics regardless of the distribution of edge weights, or the binary distribution of edges.

With a full control set $B = I_n$, the controllability Gramian can be written as,
\begin{equation} \label{eq:gramian_series}
    {W_C}(T) = \sum_{t = 0}^{T - 1} A^tBB^{\top}(A^{\top})^{t} = \sum_{t = 0}^{T - 1} A^{2t} = I + A^2 + A^4 + \cdots.
\end{equation} 
Further, in a weighted adjacency matrix $A$, the entry in the $i$-th row and $j$-th column of $A^n$ represents the strength of closed walks from node $j$ to node $i$ along paths of length $n$. \emph{Subgraph centrality} (SC) is a measure of centrality defined for unweighted networks that incorporates higher-order path lengths through a factorial discounted sum of the powers of the adjacency matrix \cite{estrada_05}. We extend the definition of subgraph centrality to a weighted adjacency matrix $A$ in order to compute the \emph{weighted subgraph centrality} as follows:
\begin{equation} \label{eq:WSC_1}
    WSC(i)=\sum_{k=0}^{\infty} \frac{(A^k)_{ii}}{k !} = 1 + (A)_{ii} + \frac{(A^2)_{ii}}{2!} + \frac{(A^3)_{ii}}{3!} + \frac{(A^4)_{ii}}{4!} + \cdots.
\end{equation}
We note that Equation \ref{eq:WSC_1} can also be written in terms of the eigenvalues and eigenvectors of $A$ \cite{estrada_05}.
\begin{equation} \label{eq:WSC_2}
    WSC(i)=\sum_{k=0}^{\infty} \frac{(A^k)_{ii}}{k !} = \sum_{k=0}^{\infty}\left(\sum_{j=1}^{N} \frac{\lambda_{j}^{k}\left(v_{j}^{i}\right)^{2}}{k !}\right),
\end{equation}
where $N$ is the number of network nodes, and $\lambda_j$ and $v_j$ are an eigenvalue and associated eigenvector, respectively. Practically, we compute weighted subgraph centrality by noting that the above definition is equivalent to selecting the diagonal entries of the matrix exponential of $A$, $WSC(i)=[\mathrm{expm}(A)]_{ii}$. Since minimum control energy and average controllability are explicitly defined in terms of the controllability Gramian, and since modal controllability is defined explicitly in terms of the network spectrum, Equations \ref{eq:gramian_series}, \ref{eq:WSC_1}, and \ref{eq:WSC_2} suggest that the weighted variant of subgraph centrality is a promising node level predictor of measures of network controllability. Hence, in the results that follow, we compute weighted subgraph centrality on the weighted adjacency matrix $A$.

\begin{figure}[!htbp]
\centerline{\includegraphics[width=\textwidth]{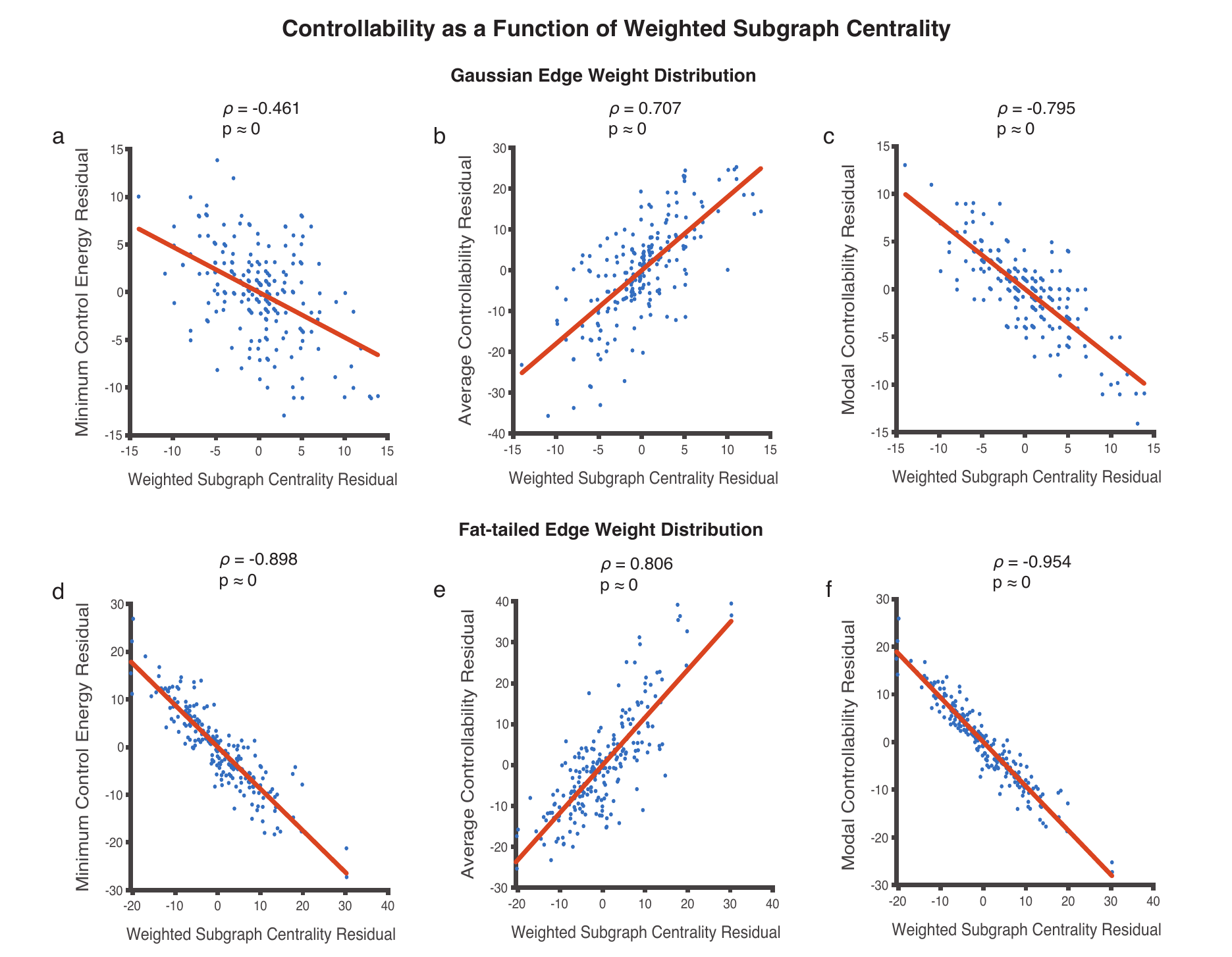}}
\caption{\textbf{Relationships between metrics of regional controllability and weighted subgraph centrality for networks approximating normal and fat-tailed distributions of edge weights.} (\textbf{a, b, c}) Weighted subgraph centrality is related in a statistically significant manner to controllability when controlling for node strength in networks with normally distributed edge weights. (\textbf{a, c}) It relates negatively with minimum control energy ($\rho = -0.461$, $p \approx 0$) and modal controllability ($\rho = -0.795$, $p \approx 0$), and (\textbf{b}) positively with average controllability ($\rho = 0.707$, $p \approx 0$). (\textbf{d, e, f}) Weighted subgraph centrality is also related in a statistically significant manner to controllability when controlling for node strength in networks with a fat-tailed distribution of edge weights. The relationships follow similar trends as networks with normally distributed edge weights; (\textbf{d}) negative with minimum control energy ($\rho = -0.898$, $p \approx 0$) and (\textbf{f}) modal controllability ($\rho = -0.954$, $p \approx 0$), and positive with (\textbf{f}) average controllability ($\rho = 0.806$, $p \approx 0$). Each point in the scatter plots represents the mean value of a controllability measure and weighted subgraph centrality across 24 (8 subjects in triplicate) network instantiations for a single brain region resulting in 234 data points. \label{fig:corr_sc}}
\end{figure}

We test weighted subgraph centrality to examine whether it is an accurate predictor of controllability that generalizes across structural brain network data sets with distinct edge weight distributions. Initially we note that weighted subgraph centrality is related negatively with minimum control energy ($\rho = -0.998$, $p \approx 0$) and modal controllability ($\rho = -0.999$, $p \approx 0$), and positively with average controllability ($\rho = 0.992$, $p \approx 0$) for Data Set 1, in which the edge weight distribution approximates a normal distribution. However, it is also related to node strength ($\rho = 0.998$, $p \approx 0$). In order to account for the effects of node strength, we perform partial Spearman rank correlations, and find that weighted subgraph centrality continues to relate negatively with minimum control energy ($\rho = -0.461$, $p \approx 0$) (Figure \ref{fig:corr_sc}a) and modal controllability ($\rho = -0.795$, $p \approx 0$) (Figure \ref{fig:corr_sc}c), and positively with average controllability ($\rho = 0.707$, $p \approx 0$) (Figure \ref{fig:corr_sc}b).

We then repeat the analyses performed above on Data Set 2, where the distribution of edge weights approximates a fat-tailed distribution. We find that weighted subgraph centrality relates negatively with minimum control energy ($\rho = -0.999$, $p \approx 0$) and modal controllability ($\rho = -0.999$, $p \approx 0$), and positively with average controllability ($\rho = 0.994$, $p \approx 0$). Since it also relates to node strength ($\rho = 0.993$, $p \approx 0$), we examine partial Spearman correlations between weighted subgraph centrality and measures of network controllability. Similar to results with Data Set 1, we find that weighted subgraph centrality continues to predict measures of network controllability in a statistically significant manner for Data Set 2. It relates negatively with minimum control energy ($\rho = -0.898$, $p \approx 0$) (Figure \ref{fig:corr_sc}d) and modal controllability ($\rho = -0.954$, $p \approx 0$) (Figure \ref{fig:corr_sc}f), and positively with average controllability ($\rho = 0.806$, $p \approx 0$) (Figure \ref{fig:corr_sc}e). Additionally, we examine the robustness of weighted subgraph centrality in predicting controllability of potentially directed structural brain networks in the Supplement. We also examine performance in an independent high resolution data set (Data Set 3) to verify generalizability of the weighted subgraph centrality - controllability relationship.

In summary, unlike participation coefficient and intra-module strength $Z$-score, weighted subgraph centrality reliably and significantly explains measures of network controllability regardless of the distribution of network edge weights.

\section{Discussion}
The topology of structural brain networks shapes and constrains the patterns of signalling between distant neuronal populations \cite{schirner2018inferring,ritter2013virtual}. These patterns, in turn, give rise to the diverse and complex large-scale functional dynamics of the brain that underlie cognition \cite{griffa2018rich,bansal2018personalized}. In this study, we sought to probe the relationship between brain network structure and the transient communication dynamics that the topology can support at the mesoscale of network organization.

While the structure-function relationship for brain networks is of interest at all scales of network organization, recent advances in community detection techniques have made the mesoscale particularly relevant \cite{betzel_18, faskowitz_18}. Distinct motifs of mesoscale structure serve different roles in the context of communication dynamics; assortative (or modular) interactions allow for information integration and segregation \cite{park_13, goni_13}, core-periphery motifs with rich-club hubs \cite{colizza2006detecting} allow for information broadcast and receipt \cite{van_den_heuvel_12, van_den_heuvel_13}, while disassortative motifs support information transmission. Controllability, by contrast, influences state transitions \cite{towlson2018celegans}, and has been related to the notion of cognitive control, where the brain shifts from one cognitive state to another \cite{cornblath_19}. Through our numerical simulations, we demonstrate that distinct features of community structure are likely to be implicated in distinct aspects of neural computation. 

A mesoscale feature is any topological feature that cannot be explained by the local neighborhood of a node, and is better explained by larger neighborhoods around the node, than it is by the total global architecture \cite{lohse2014resolving,schlesinger2017improving}. Much of the literature has focused on modularity and core-periphery structure as the canonical forms of mesoscale structure \cite{girvan2002community, newman_04}. But our results suggest that another distinct form of mesoscale structure must be considered, and that is the feature that drives controllability statistics \cite{kim_18}. Here we demonstrate that \emph{weighted subgraph centrality}, can potentially assess this distinct dimension of mesoscale architecture in future studies.

Recent work has sought to define measures of network topology, such as disassortativity and core-peripheriness, both at the scale of nodes and at the scale of communities \cite{zhang_15, sarkar_13, sarkar_18,foster2010edge}. A natural direction to extend this work is to examine the distribution of eigenvalues as the network topology gradually alters to become more assortative or core-periphery from disassortative. Moments of the eigenvalue distribution such as the mean, variance, skewness, and kurtosis may hold valuable insights into the behavior of network control metrics as functions of mesoscale architecture and edge-weight distribution. More theoretical work is needed in order to relate the spectra of weighted graphs to properties of network controllability. Recent work has attempted to create closed-form characterizations of spectral properties for both assortative \cite{van_mieghem_10} and core-periphery networks. In addition, since structural brain networks simultaneously possess a variety of community interaction motifs \cite{betzel_18}, future work might involve characterizing the effects of mixed interactions in numerical simulations similar to those performed in this work.

Controllability statistics cannot be explained simply by node strength, nor can they be explained by mesoscale structure. Through our results, we verify that node strength is a significant predictor of network controllability in the classes of graphs we study. However, it does not uniquely explain controllability. In all our analyses, after verifying the dependence of controllability on node strength, we proceed to regress out its effects when examining any dependence on other metrics of interest. We demonstrate in the Supplement that weighted subgraph centrality correlates more strongly, as well as linearly, with measures of network controllability than node strength does across a range of values of the time horizon of control. Additionally, whereas weighted subgraph centrality survives corrections for node strength, and continues to significantly predict controllability, modularity often does not.  This distinction indicates that weighted subgraph centrality explains parts of network controllability that neither node strength nor any modularity metric we evaluated are able to. 

Our results indicate that higher-order path-dependent network structure, as captured by weighted subgraph centrality, is strongly related to transient communication dynamics. Indeed, it explains controllability better than descriptive statistics such as node strength and measures of modularity. At the network-level communicability is able to separate patients of stroke from healthy controls \cite{crofts_11}. Communicability metrics have been shown to be sensitive indicators of lesions in patients with relapsing-remitting multiple sclerosis \cite{li_RRMS}. It has also been shown that communicability is disrupted in patients of Alzheimer's disease \cite{lella_18}. Weighted subgraph centrality is the weighted extension of the notion of self-communicability. The consistently strong relationship between weighted subgraph centrality and measures of network controllability, suggests that statistics derived from linear control theory (such as average and modal controllability, and minimum energy) are also likely useful tools in investigating the disruptions to brain network dynamics in disease.

The distinction between modularity and controllability impacts our interpretation of previous reports that provide evidence that these two features change appreciably over normative neurodevelopment. A naive hypothesis could be that the change in modularity drives a change in controllability, or \emph{vice versa}. However, Tang \emph{et al.} show that their network controllability results hold after regressing out modularity \cite{tang_17}. Moreover, we find more generally using multiple data sets and systematic variation of network modularity in simulations, that the two variables cannot be explained by one another. In the context of development, our results suggest that the process of brain development may reflect a more complex optimization function that coordinates a change in modularity alongside a change in controllability. What that function is, and what the mechanism of coordination is, remains to be clearly specified, but would be an important area for future work. The distinction between modularity and controllability also calls for care when interpreting reports of either of these features changing as a function of aging \cite{baum_17}, training \cite{arnemann2015functional}, treatment \cite{baliki2018brain,tao2019effects}, injury \cite{gratton2012focal}, or disease \cite{vertes2012simple}.

\subsection{Biophysical interpretation of model parameters}
In the discrete-time LTI framework, the variable $\bm{x}(t)$ is a real $N$-dimensional vector, whose $i$-th element corresponds to the level of activity of brain region $i$. The level of activity of each brain region can be defined in multiple ways, such as the average blood oxygen level dependent (BOLD) signal from functional magnetic resonance imaging (fMRI) \cite{cui_20,braun2019brain}, or the average electrical activity from electrophysiological recordings \cite{stiso_19,khambhati2019functional}. As for the inputs, the variable $\bm{u}(t)$ represents independent control inputs whose influence can be linearly separated from the activity along white matter tracts. For instance, these influences may be endogenous neurotransmitter activity \cite{braun2019brain}, task-based internal modulation of the brain state \cite{cornblath_18,cui_20}, or exogenous inputs such as pharmacological agents \cite{braun2019brain}, direct electrical stimulation or transcranial magnetic stimulation \cite{stiso_19,khambhati2019functional}.

Hence, while the most immediate and straightforward interpretation of $\bm{u}(t)$ is as an external electrical or pharmacological perturbation, we do not discount the possibility of other internal neural mechanisms (e.g., local dynamics of gray-matter neurons) that are independent of and take advantage of these white-matter tracts to influence global dynamics. Keeping both possibilities in mind, we refer to $\bm{u}(t)$ as the ``exogenous input'' for conceptual tractability. In addition, if it is easier for an exogenous input to globally influence the system by changing the activity of a node (less energetic cost, more spread of activity), then it is similarly easier for the endogenous activity of that node to globally influence the system. If the endogenous nodal activity is generated by a process that is independent of the white-matter tracts, it can be modeled as a separate input $\bm{u}(t)$ to the linear dynamical system without making additional assumptions beyond an interpretation of exogenous inputs.

In the context of structural brain networks and computations of control energy for state transitions, more work is needed to neurobiologically motivate the choices for initial and target states. Prior work has made imaging-based choices for states to model cognitive states of the brain, such as band-limited power \cite{stiso_19} or beta weights from a general linear model of BOLD activation from functional magnetic resonance imaging \cite{braun2019brain}. Alternatively, binary activation of regions corresponding to functional modules has also been examined \cite{betzel_16}. However, since the focus of this paper is to examine network controllability from the perspective of network community structure, a thorough investigation of state-pair choices is beyond the current scope. Our specific choice here is motivated by prior work probing the generic control properties of a system by formulating an influence maximization problem \cite{NIPS2016_6119}. We compute minimum control energies by performing $N$ state transitions to $N$ one-hot vectors for each brain region $i$, such that the energies $E_i$ form an upper bound on the energy required to perform arbitrary non-negative state transitions $\bm{x}^*$ (see Supplement for more discussion).

\subsection{Methodological considerations}
The choice of the weighted stochastic block model (WSBM) to uncover network communities is motivated by the desire to uncover community interaction motifs extending beyond the traditionally examined assortative type. We hypothesized that disruptions to specific motifs ought to result in motif-specific profiles of network controllability. In the context of empirical brain data, the WSBM uncovers a diverse community structure reflecting the diversity of the functional dynamics supported. The WSBM is an incredibly flexible community detection technique. However, this flexibility comes at the price of having to choose a number of parameters \textit{a priori}, including the number of communities that are anticipated to exist in the network, and a prior regarding the nature of the edge weight distribution. We fix the number of communities by sweeping over a range of values and choosing the value that maximizes the likelihood of observing the given network data. Additionally, we verify salient analyses performed in the paper in the Supplement with a different choice of edge weight distribution prior.

In our network-level numerical simulations, we adopt the geometric distribution as a representative fat-tailed distribution from which to draw edge weights. The geometric distribution is the discrete counterpart to the exponential distribution. Another fat-tailed distribution that is commonly explored in network neuroscience is the scale-free distribution characterized by a power-law \cite{wu_yan_18, sizemore_16}. However, recent work has demonstrated that scale-free networks are not as ubiquitous as previously thought, and that the exponential distribution is often a suitable alternative \cite{broido_19}. Our motivation in considering the normal and geometric distributions was to examine controllability of networks with two different edge weight distributions. Future work could characterize controllability performance explicitly for networks with a scale-free distribution of edge weights, instead of relying on a stand-in fat-tailed distribution \cite{wu_yan_18}.

While a linear model of network dynamics lends itself well to control-theoretic studies of communication dynamics, empirical results have shown that brain activity is non-linear \cite{rabinovich_06}. However, recent work has demonstrated that a linear approximation is often useful \cite{honey_09, galan_08, muldoon_16}. In addition, the linear framework can be adapted to incorporate more complex features of neural dynamics \cite{yang_19, li_17, zanudo_17}. Similar to the WSBM, applying linear network control theory to empirical data involves setting a variety of hyper-parameters, such as the time horizon over which control is exerted, the target state vector in computations of minimum control energy, or the normalization scheme employed. Our hyper-parameter choices are motivated by the desire to investigate and compare network topology across data sets with very distinct edge weight distributions. As a result, we choose a non-zero short time horizon after scaling down the network adjacency matrices by their largest eigenvalues. This step ensures that the fastest evolving modes across systems stay consistent. However, we note the need for further work to motivate parameter choices from a neurophysiological perspective.

Our results demonstrate that the choice of empirical measurement that is used to characterize structural edges in brain networks is crucial to investigations of network control. For instance, whereas results derived from quantitative anisotropy (QA) weighted networks may lead us to conclude that modularity as measured by the participation coefficient and average controllability are related (Figure \ref{fig:corr_gauss}), streamline count weighted networks present contrary results (Figure \ref{fig:corr_stream}). It is unclear if one type of empirical estimate for network edges in structural brain networks is better than another. It is possible that some measures better assess signal speed, others better assess bundle volume, and yet others better assess micro-structure integrity \cite{johansen2010behavioural}. Perhaps the choice of edge weight definition also has implications for community detection. For instance, are network partitions likely to be different depending on the distribution of edge weights? More work is needed to contextualize the impact of edge weights on our interpretations of modularity, core-periphery structure, and network controllability, and their relationships to communication, computation, and dynamics. The WSBM continues to remain a promising tool in this endeavor since it is comprised of a generative model with a prior over the edge weight distribution built into its framework.

\section{Conclusion}
We began with the hypothesis that the extent of a node's participation in the network community structure ought to be related to its controllability. We find that modularity as measured by the participation coefficient and intra-module strength Z-score is a significant predictor of average and modal controllability for structural brain networks where the distribution of edge weights approximates a normal distribution. For these networks, neither participation coefficient nor intra-module strength Z-score are related with minimum control energy. For networks where edge weights approximate a fat-tailed distribution, we find that participation coefficient is not related in a statistically significant manner with any measure of network controllability. Intra-module strength Z-score is related with minimum control energy and average controllability, but not with modal controllability.

By contrast, \emph{weighted subgraph centrality} is a statistically robust predictor of network controllability, regardless of the distribution of network edge weights. The relationships between weighted subgraph centrality and measures of network controllability, indicate that higher-order path-dependent network structure predicts transient communication dynamics. At the network level, through numerical simulations, we demonstrate that binary topology alone is not a predictor of mean network controllability. Along a structural continuum from disassortative to assortative, or from disassortative to core, mean controllability profiles are heavily dependent on the distribution of network edge weights. Our study contributes to an understanding of how the diverse mesoscale structural architecture of the brain, characterized by a variety of community interaction motifs and edge weight distributions, supports transient dynamics in the brain. 

\section{Methods}

\subsection{Data}
Structural brain networks used in the analyses are constructed from diffusion spectrum imaging (DSI) data acquired in triplicate from eight subjects (mean age $27 \pm 5$ years, two female, two left handed) along with T1-weighted anatomical scans at each scanning session. DSI scans sampled $257$ directions using a Q5 half-shell acquisition scheme with a maximum $b$-value of $5000$ and an isotropic voxel size of $2.4$ mm. Axial acquisition with the following parameters was employed: repetition time (TR) = $11.4$ s, echo time (TE) = $138$ ms, $51$ slices, field of view (FoV) ($231$, $231$, $123$ mm). All participants volunteered with informed consent in accordance with the Institutional Review Board/Human Subjects Committee, University of California, Santa Barbara. Data acquisition and network construction methods are described elsewhere in further detail \cite{gu_15}.

The data contain brain networks where edges represent diverse estimates of inter-node connections, including white matter streamline counts between regions, mean quantitative anistropy (QA) values along the streamlines, and generalized fractional anisotropy (GFA) values. The choice of edge definition has implications for the distribution of edge weights in the networks. Streamline counts have a fat-tailed edge weight distribution, whereas QA values are normally distributed. In the present study, we investigate the implications of edge weight distribution on network controllability by using networks with streamline counts as well as QA values. We refer to networks with QA values as Data Set 1, and to networks with streamline counts as Data Set 2.

Additionally, we repeat salient analyses in the Supplement on a higher resolution data set, henceforth termed Data Set 3. This data set is acquired from ten healthy human subjects as part of an ongoing data collection effort at the University of Pennsylvania; the subjects provided informed consent in writing, in accordance with the Institutional Review Board of the University of Pennsylvania. Similar to Data Set 2, Data Set 3 is comprised of structural brain networks where edges reflect streamlines counts between regions.

For Data Set 3, all scans are acquired on a Siemens Magnetom Prisma 3 Tesla scanner with a $64$-channel head/neck array at the University of Pennsylvania. All participants volunteered with informed consent in accordance with the Institutional Review Board/Human Subjects Committee, University of Pennsylvania. Each data acquisition session includes both a diffusion spectrum imaging (DSI) scan as well as a high-resolution T1-weighted anatomical scan. The diffusion scan is $730$-directional with a maximum $b$-value of $5010$s/mm2 and TE/TR = $102$/$4300$ ms, which includes $21$ $b = 0$ images. Matrix size is $144 \times 144$ with a slice number of $87$. Field of view is $260 \times 260$ mm$^2$ and slice thickness is $1.80$ mm. Acquisition time per DTI scan is $53:24$ min, using a multiband acceleration factor of $3$. The anatomical scan is a high-resolution three-dimensional T1-weighted sagittal whole-brain image using a magnetization prepared rapid acquisition gradient-echo (MPRAGE) sequence. It is acquired with TR = $2500$ ms; TE = $2.18$ ms; flip angle $= 7$ degrees; $208$ slices; $0.9$ mm thickness. More detail on data acquisition and processing is available elsewhere \cite{kim_18}.

\subsection{Weighted Stochastic Block Model}
In our effort to probe the relationship between network controllability and the mesoscale architecture of structural brain networks, the first step is to partition the networks into communities. We apply block modeling to infer network partitions from data. Block models uncover diverse mesoscale architectures \cite{hastings_06, aicher_15}, which may have implications for network controllability. The model assumes that connections between nodes are made independently of one another, and that the probability of a connection between two nodes depends only on the communities to which the nodes are assigned. Fitting the model involves estimating the parameters that maximize the likelihood of observing a given network. 

The Stochastic Block Model (SBM) seeks to partition the nodes of a network into $K$ communities. Let $z_i \in \{1,\cdots,K\}$ indicate the community label of node $i$. Under the block model, the probability $P_{ij} = \theta_{z_i, z_j}$ that any two nodes $i$ and $j$ are connected depends only on their community labels, $z_i$ and $z_j$, where ${z_i, z_j} \in \{1,\cdots,K\}$. To fit the block model to the observed data in $A$, we estimate $\theta_{rs}$ for all pairs of communities $\{r,s\} \in \{1,\cdots,K\}$ and the community labels $z_i$. Assuming that the placement of edges is independent of one another, the likelihood of the SBM having generated a network is
\begin{equation} \label{eq:SBM}
P(A \mid \{z_i\}, \{\theta_{rs}\}) = \prod_{i,j} ({\theta_{z_iz_j}})^{A_{ij}}({1-\theta_{z_iz_j}})^{1-A_{ij}}.
\end{equation}
Fitting the SBM involves determining the parameters $\{z_i\}$ and $\{\theta_{rs}\}$. However, the SBM is limited to binary networks. By contrast, the weighted stochastic block model (WBSM) \cite{hastings_06, aicher_13, aicher_15} incorporates edge weights into its framework making weighted graphs such as brain networks accessible to block models for community detection \cite{pavlovic_14, betzel_18, faskowitz_18, faskowitz_19}.

In the weighted variant (WSBM) of the block model, the likelihood function in Eq.~\eqref{eq:SBM} is modified to
\begin{equation} \label{eq:WSBM}
P(A \mid \{z_i\}, \{\theta_{rs}\}) \propto exp\bigg(\sum_{i,j}T(A_{ij})\;.\;\eta(\theta_{z_iz_j})\bigg).
\end{equation}
In the binary case (SBM), $T$ and $\eta$ correspond to the vector-valued function of sufficient statistics and the vector-valued function of natural parameters for the Bernoulli distribution, respectively. Different choices of $T$ and $\eta$ can allow for the edge weights to be drawn from different distributions of the exponential family. The WSBM, just like its classical variant, is parameterized by the set of community assignments, $\{z_i\}$, and the parameters $\{\theta_{rs}\}$. The difference is that each $\theta_{z_iz_j}$ now specifies the parameters governing the weight distribution of the edge $z_iz_j$, and not the probability of edge existence. For the normal distribution, the vector-valued function of sufficient statistics is $T = [x, x^2, 1]$, while the vector-valued function of natural parameters is $\eta = [\mu/\sigma^2, -1/2\sigma^2, \mu^2/(2\sigma)^2]$. Edges are now parameterized by a mean and variance, $\theta_{z_iz_j} = (\mu_{z_iz_j}, {\sigma^2}_{z_iz_j})$. As a result, the likelihood function in Eq.~\eqref{eq:SBM} can be modified to read 
\begin{equation} \label{eq:normal_WSBM}
    P\big(A \mid \{z_i\}, \{\mu_{rs}\}, \{{\sigma^2}_{rs}\}\big) = \prod_{i,j} exp\bigg(A_{ij}\cdot\frac{\mu_{z_i,z_j}}{\sigma^2_{z_iz_j}} - A_{ij}^2\cdot\frac{1}{2\sigma^2_{z_iz_j}} - 1\cdot\frac{\mu^2_{z_i,z_j}}{\sigma^2_{z_iz_j}}   \bigg)
\end{equation}
for edge weights drawn from the normal distribution.

An additional challenge in fitting block models to data is the handling of sparse networks \cite{aicher_15}. This is particularly important for brain networks since the neural connectome is sparse and most entries in the adjacency matrix $A$ are zero. This sparsity is handled by modeling edge weights as described above, and separately modeling edge presence with a Bernoulli distribution. If $T_e$ and $\eta_e$ represent the edge existence distribution, and $T_w$ and $\eta_w$ the edge weight distribution, the likelihood function for $A$, can be written as:
\begin{equation} \label{eq:sparse_WSBM}
    log P(A \mid \{z_i\}, \{\theta_{rs}\}) = \alpha\sum_{i,j \in E}T_e(A_{ij})\;.\;\eta_e(\theta_{z_iz_j}) + (1-\alpha)\sum_{i,j \in W}T_w(A_{ij})\;.\;\eta_w(\theta_{z_iz_j}).
\end{equation}
In Eq.~\eqref{eq:sparse_WSBM}, $E$ is the set of all edges and $W$ is a subset of $E$ representing the weighted edges. A variational Bayes algorithm is then used to estimate the model parameters from data, as outlined in \cite{aicher_13} and \cite{aicher_15}.

However, this pipeline is still incomplete as fitting the weighted stochastic block model (WSBM) to a network requires that the number of blocks $K$ in the community structure be chosen \textit{a priori}. A data-driven approach can help determine the suitable number of blocks present. Since the WSBM is a generative model, we can estimate the likelihood (see \eqref{eq:sparse_WSBM}) of observing a connectivity matrix $A$ for different values of $K$. The $K$ that maximizes the likelihood of observing the data is chosen as the parameter value when inferring network partitions downstream. For Data Set 1 and Data Set 2, we run the WSBM on all structural connectivity matrices derived from the eight subjects ($8$ subjects $\times \; 3 = 24$ matrices) while sweeping over a range of $K$ values from $K = 6$ to $K = 15$. Since the WSBM is not deterministic, we run $10$ iterations for each subject for each trial at each choice of $K$. We find that data likelihood is maximized when $K = 12$ for networks with normally distributed edge weights (Data Set 1) with a Gaussian edge weight prior, and when $K = 14$ for networks with a fat-tailed edge weight distribution (Data Set 2) with a log-normal edge weight prior. A by-product of the process of selecting $K$ is the partitions of the networks into communities that we seek. At the $K$ that maximizes data likelihood, each network already has $10$ instantiations of partitions. The network partition chosen for the analyses is the one that is the most central out of all, as defined by variation of information \cite{faskowitz_18}. For Data Set 3, we run $25$ iterations of the WSBM for each $K$ and find that the likelihood is maximized when $K = 10$ with a log-normal edge weight distribution prior.

Code to infer community structure from networks using the WSBM is freely available at \url{http://tuvalu.santafe.edu/~aaronc/wsbm/} \cite{aicher_13, aicher_15}.

\subsection{Network Statistics}
Recall that our hypotheses depend on the quantification of the extent to which nodes participate in interactions with nodes from other communities. We compute the participation coefficient \cite{guimera_05}, and intra-module strength $Z$-score \cite{guimera_05} to quantify this extent based on the WSBM-generated partitions of brain networks. 

The participation coefficient for a node $i$ is defined as
\begin{equation} \label{eq:part_coeff}
    PC_i = 1 - \sum_{z = 1}^{K} \big(\frac{\kappa_{iz}}{\kappa_i}\big)^2,
\end{equation} 
where $\kappa_{iz}$ is the strength of connection of node $i$ to nodes in community $z$, and $\kappa_i$ is the total strength of node $i$. The term $K$ is the number of communities in the partition. Intra-module strength $Z$-score ($Z$) for node $i$ is defined as
\begin{equation} \label{eq:z_score}
    Z_i = \frac{\kappa_{iz_i} - \bar{\kappa_{z_i}}}{\sigma_{\kappa_{z_i}}},
\end{equation}
where $\kappa_{iz_i}$ is the strength of connection of node $i$ to other nodes in its own community $z_i$, $\bar{\kappa_{z_i}}$ is the average strength of connection of all nodes in module $z_i$ to other nodes in $z_i$, and $\sigma_{\kappa_{z_i}}$ is the standard deviation of $\kappa_{iz_i}$. We compute these metrics using freely available code from the \texttt{Brain Connectivity Toolbox} (\url{https://sites.google.com/site/bctnet/}) \cite{rubinov_10}.

At the network level, the modularity quality index $Q$ measures how well a given partition of a network compartmentalizes its nodes into modules \cite{newman_04, newman_06}. We use this measure in conjunction with numerical simulations to quantify the extent of modularity at the network level. $Q$ is defined as:
\begin{equation}
    Q=\sum_{i j}\left[A_{i j} - N_{i j}\right] \delta\left(z_{i}, z_{j}\right),
\end{equation}
where $N_{ij}$ is the expected strength of connections between nodes $i$ and $j$ under the Newman-Girvan null model, which is designed to quantify assortativity \cite{newman_06}. The Kronecker delta function equals $1$ when the two nodes belong to the same community, and equals zero otherwise. 

\subsection{Numerical Simulations\label{sec:simulations}} 
In order to generate networks with specific edge weight distributions and binary topologies, we make use of a $2 \times 2$ block structure, and specify the binary density of each block separately. When the fraction of total edges inside of the on-diagonal blocks exceeds the fraction in the off-diagonal blocks, the network has an assortative community structure. By contrast, when the fraction of total edges in the off-diagonal blocks exceeds the fraction inside of the diagonal blocks, the network has a disassortative community structure. If the fraction of edges inside of the block in the $[1, 1]$ position is higher than the fractions for the three remaining blocks, the network has a core-periphery architecture. Upon fixing the value of the fraction of total edges inside of a block of interest, the remaining edges are distributed across the network such that the network's binary density remains $0.1485$, which is the mean density of structural brain networks from Data Set 1.

For each edge, a corresponding weight value is drawn from a pre-specified distribution, either a normal distribution or a family of geometric distributions (see below). Edges drawn from the normal distribution are parameterized by $\mu = 0.5$ and $\sigma = 0.12$ \cite{wu_yan_18}. The geometric distribution was chosen as a representative of the family of fat-tailed distributions that are ubiquitous in biological systems \cite{wu_yan_18, sizemore_16, broido_19}. Geometric distributions are parameterized by a single number $p$, which represents the probability of success of a Bernoulli trial. Weights are then assigned to edges by incrementing the value of an edge until the first failure of a Bernoulli trial. Therefore, when $p$ is closer to $0$ edge weights tend to remain small, and when $p$ is closer to $1$ edge weights tend to take on large values.

During the course of numerical simulations along a structural continuum from disassortative to assortative, or from disassortative to core-periphery, new networks are created at each stage with new binary densities for the four blocks. In the case of the continuum from disassortative to assortative networks, the fraction of total edges in the on-diagonal blocks is gradually altered. When this fraction is $0$, all network edges lie in the off-diagonal blocks giving the network a disassortative architecture. By contrast, when the fraction is $1$ and all edges lie inside of the on-diagonal blocks, the network is perfectly modular and possesses an assortative mesoscale structure. In the case of the continuum from disassortative to core-periphery networks, the fraction inside of the $[1, 1]$-block is gradually altered, and the $[2, 2]$-block is left empty. When the fraction of total edges inside of the $[1, 1]$-block is $0$, the network is disassortative, whereas when the fraction is $1$, the network only has a single densely connected core. Alternatively, this process may be thought of as moving edges from the off-diagonal blocks to either the on-diagonal blocks, or the $[1, 1]$-block, depending on the structural continuum under consideration.

At each stage along the continuum, $100$ networks are created using the set of parameters that define the network topology of the ensemble. The process of creating ensembles is intended to ensure roughly similar degree distributions for networks across a structural continuum. In case of simulations for networks with geometrically distributed edge weights, a further constraint is enforced. In order to align network topology to the network geometry, when drawing edge weights for the numerical simulations, we use multiple geometric distributions. For each block in the $2 \times 2$ block adjacency matrix, $p$ is chosen to be the desired binary density (fraction of total edges) corresponding to the block \cite{wu_yan_18}. We summarize the extent of modularity in each network in an ensemble along the continuum using the modularity quality index $Q$. Since networks are generated with partitions that are known \textit{a priori}, we do not perform a re-partitioning of the networks in order to determine $Q$. We characterize the relationship between $Q$, and the fraction of edges inside of modules (as well as inside the core) in the Supplement.

\newpage
\section{Citation Diversity Statement}
Recent work in neuroscience and other fields has identified a bias in citation practices such that papers from women and other minorities are under-cited relative to the number of such papers in the field \cite{dworkin_20, maliniak_13, caplar_17, chakravartty_18, thiem_18}. Here we sought to proactively consider choosing references that reflect the diversity of the field in thought, form of contribution, gender, race, geography, and other factors. We used automatic classification of gender based on the first names of the first and last authors \cite{dworkin_20}, with code freely available at \url{https://github.com/dalejn/cleanBib}. Possible combinations for the first and senior authors include male/male, male/female, female/male, and female/female. After excluding self-citations to the first and senior authors of our current paper, the references in this work contain 58.6\% male/male, 8\% male/female, 18.4\% female/male, 3.4\% female/female, and 11.5\% unknown citation categorizations. We look forward to future work that could help us better understand how to support equitable practices in science.

\section{Acknowledgments}
The authors gratefully acknowledge helpful discussions with Jennifer Stiso, Dr. Eli J. Cornblath, Dr. Xiasong He, and Dr. Ann Sizemore-Blevins. DSB would like to acknowledge support from the John D. and Catherine T. MacArthur Foundation, the Alfred P. Sloan Foundation, the ISI Foundation, the Paul Allen Foundation, the Army Research Laboratory (W911NF-10-2-0022), the Army Research Office (Bassett-W911NF-14-1-0679, Grafton-W911NF-16-1-0474, DCIST- W911NF-17-2-0181), the Office of Naval Research, the National Institute of Mental Health (2-R01-DC-009209-11, R01 MH112847, R01-MH107235, R21-M MH-106799), the National Institute of Child Health and Human Development (1R01HD086888-01), National Institute of Neurological Disorders and Stroke (R01 NS099348), and the National Science Foundation (BCS-1441502, BCS-1430087, NSF PHY-1554488 and BCS-1631550).

\section{Author Contributions} 
S.P.P. performed the simulations, analyzed the data, made the figures, and wrote the paper. J.Z.K. contributed analytical solutions. J.Z.K., F.P., and D.S.B. participated in discussions and edited the paper.

\newpage
\bibliographystyle{unsrt}
\bibliography{./bibfile.bib}

\end{document}